# Hybrid Thermal Yagi-Uda Nanoantennas for Directional and Narrow Band Long-wavelength IR Radiation Sources.

Marco Centini[1,2], Maria Cristina Larciprete[1], Roberto Li Voti[1], Mario Bertolotti[1], Concita Sibilia[1] and Mauro Antezza[1,2,3].

[1] Sapienza University of Rome Department of Basic and Applied Sciences for Engineering, Via A. Scarpa 14, I-00161 Rome, Italy

[2] Laboratoire Charles Coulomb (L2C), UMR 5221 CNRS-Université de Montpellier, F-34095 Montpellier, France

[3] Institut Universitaire de France, 1 rue Descartes, F-75231 Paris, France

Abstract
We investigate the possibility of spatially and spectrally controlling the thermal infrared emission by exploitation of the Yagi–Uda antenna design. Hybrid antennas composed of both SiC and Au rods are considered and the contributions of emission from all the elements, at a given equilibrium temperature, are taken into account. We show that the detrimental effect due to thermal emission from the not ideal parasitic elements drastically affect the performances of conventional thermal Au antennas in the 12 µm wavelength range. Nevertheless, our results show that the hybrid approach allows the development of efficient narrow-band and high directivity sources. The possibility of exploiting the Yagi-Uda design both in transmission and in reception modes, may open the way to the realization of miniaturized, efficient, robust and cheap sensor devices for mass-market applications.

Introduction:

The availability of low cost, integrated, radiation sources in the infrared range with narrow-band emission spectrum and good directional selectivity is of great importance in a variety of applications such as infrared sensing [1], thermophotovoltaics [2], radiation cooling [3], and thermal circuits [4]. With particular reference to gas sensing applications, narrow band emission sources with a good level of spatial coherence are required in order to optimize the signal to noise ratio and to increase selectivity and sensitivity. Quantum cascade lasers (QCL) [5] are probably the best choice for mid to far IR spectroscopy due to their high spectral density and single-mode operation. Using a set of QCL tuned at different wavelengths it could be possible to match individual absorption lines of different chemical species, for example. Nevertheless, the practical application of these sources for commercial sensing devices for home, automotive markets as well as smartphone/laptop accessories are strongly limited by high fabrication costs.
A common and inexpensive way to obtain infrared radiation is to take advantage of thermal emission from a heated body. However, the spectral and directional control of thermally emitted light is a challenging task due to its incoherent behavior (both spatially and temporally). In the last decade several approaches have been proposed to control the mechanisms of thermal radiation and heat exchange on the nanoscale [6-8] also in 2D and multilayered materials [9-14]. Enhancement and control of thermal radiation have been investigated in gratings [15], metamaterials [16-19], metasurfaces [20] and single nanoantennas [21]. It has been shown, both theoretically and experimentally, that sub-wavelength patterning of materials supporting surface polaritons results in narrow emission lobes. Enabling the exploitation of near field interactions between different objects and the resonant coupling of evanescent modes into propagative ones, it is possible to obtain efficient and partially coherent emission. The mechanism at the base of these tailored emission is the excitation of resonant surface polaritons. Metals have been extensively studied in the visible and near IR range; enhancement and selectivity of thermal emission have been obtained thanks to resonant plasmon polariton excitation. Typically, these effects are stronger in a spectral range close to the plasma frequency of the metal and become weaker in the mid-wavelength IR (MWIR). For efficient sources in the long-wavelength IR (LWIR, 8-15 µm) it is possible to rely on polar materials supporting phonon polaritons. Thermal emission from silicon carbide (SiC) gratings or silica ($SiO_2$) patterned structures has been intensively studied [22,23]. Moreover the emission properties of a single SiC rod antenna have been studied by numerical calculations and experiments [24]. Recently it has also been shown that it is possible to use porous SiC to create narrow-band emitting/absorbing metamaterials [25] with a certain degree of tunability in order to match absorption lines of hazardous materials [26,27] in the 12-14 µm wavelength range.

For sensing applications, the growing requirement for device miniaturization and the demand for precisely controllable and tunable sources made the nanoantenna based approaches very appealing. In the optical and near infrared ranges remarkable results have been obtained by using gold Yagi-Uda (YU) nanoantennas coupled with quantum dots emitters [28]. The YU design consists on a feeder which plays the role of the emitter, a reflector which is used to suppress backward radiation and a set of directors which are responsible for the directivity of the emission pattern as sketched in figure 1. This scheme is widely used for radiofrequency waves; The feeder is typically a half wavelength dipole emitter and it is the only element that is electrically excited. The parasitic elements (reflector and directors) are also half wavelength dipole antennas but with a shorter (directors) or longer (reflector) resonant wavelength in order to modify the phase of the elements' current with respect to its excitation from the feeder element. The enhanced directionality and backward suppression are achieved by selective control in forward and backward direction of destructive and constructive interference between emitted and scattered waves. This is obtained by placing the elements at the proper distances from each other. The efficiency of the device is related to the coherence of the signal emitted by the feeder and by the low losses induced by the parasitic elements. For example, the performance of the optical YU antenna presented in Ref. 28 relies on the narrow spectral line emission from the quantum dot that resonantly couples to the feeder and to the fact that all the other elements are passive.

Recently some possible scheme of metallic antennas and composite antennas for mid-IR range have been proposed [29] and experimentally investigated [30,31]. Concerning metallic antennas, it must be noted that the field enhancement and spectral selectivity of a resonant antenna drastically decreases at longer wavelengths. Thus, the broad band thermal radiation originating from chaotic fluctuations is responsible for highly incoherent signals which dramatically reduce the effect of interference due to the lack of coherence. In Ref.32 a gold YU design has been investigated for emission in the wavelength range around 7 µm. Only the emission related to the heated feeder is considered while the other elements are supposed to act as passive scatterers. The authors report a limited effect in directionality due to the fact that the frequency spectrum at the resonance peak is quite large ($\Delta\lambda$=1 µm with respect to the peak of emission at $\lambda$=7.25 µm).
Moreover, a realistic evaluation should consider also the emission contributions from the reflector and the directors that are, reasonably, at the same temperature of the feeder.

Indeed, the directors (reflector) are antennas as well, with respectively a smaller (larger) resonant wavelengths. Thus they will contribute to broaden the overall spectrum of emission. Following this argumentation, we will show that a gold thermal YU antenna has a limited efficiency in LWIR range (around $\lambda$=12 µm) and its behavior is mostly driven by the emission of the directors rather than the feeder.

In this manuscript we propose a Hybrid thermal YU antenna composed of gold and SiC, operating in the LWIR range. In particular, we consider a SiC dipole antenna as the feeder for its narrow band high emissivity and gold rods acting as parasitic elements. Our numerical results show that despite of the detrimental noise introduced by the emissivity of the metallic reflector and directors, the efficiency of the typical boosting effect of the YU design is preserved. Moreover the tolerances are extremely robust and the same antenna design can be used to tune the emission peak of the antenna in the reststrahlen band of the SiC by only changing the length of the feeder rod. These features make thermal YU hybrid antennas very appealing for use as multiple arrays of narrow band high directivity sources for different species of gas sensing.

Materials and Methods:

As a starting point we compare the thermal emission of gold and SiC dipole antennas to put into evidence the difference in the bandwidth of emission spectra.

The emitted power spectral density is evaluated by applying a previously developed model based on the fluctuational electrodynamics approach and on the discretization of the resulting volume integral equation to calculate emissivity and far field spatial emission pattern of nanoparticle ensembles [33]. More in details, the numerical model is based on the fluctuation dissipation theorem [34]: for a medium with relative permittivity $\varepsilon_r$ at a given temperature T the cross correlation spectral density for the chaotic current density related to thermal fluctuations is given by the expression:

$$\langle \vec{J}^0_\omega(\vec{r}_1) \vec{J}^{0\dagger}_{\omega'}(\vec{r}_2) \rangle = \frac{\omega}{\pi} \varepsilon_0 \text{Im}(\varepsilon_{r,\omega}) \Theta_\omega(T) \delta(\vec{r}_1 - \vec{r}_2) \delta(\omega - \omega') \bar{\bar{I}} \qquad (1)$$

Where the brackets represent a statistical ensemble average, $\delta$ is the Dirac function, $\omega$ is the angular frequency, $\bar{I}$ is the identity operator and:

$$\Theta_\omega(T) = \frac{\hbar\omega}{\exp(\hbar\omega/k_B T) - 1}; \qquad (2)$$

being $k_B$ the Boltzmann constant and $\hbar$ the reduced Planck constant. The symbol † stands for the operation of Hermitian conjugation. In Ref.33 we have shown that the cross spectral density defined as:

$$\bar{\bar{W}}_\omega(\vec{r}_1, \vec{r}_2) \delta(\omega - \omega') = \langle \vec{E}_\omega(\vec{r}_1) \vec{E}^\dagger_{\omega'}(\vec{r}_2) \rangle \qquad (3)$$

can be numerically evaluated on a discretized mesh as a product between an auxiliary deterministic matrix, taking into account for the scattering problem, and the cross correlation function of the fluctuating sources given by the fluctuation dissipation theorem (see eq. 1). In the far-field regime, the diagonal elements of the cross spectral density are proportional to the mean radiated intensity spectral density. Thus, the average power spectral density emitted by the source and collected at a given distance $d$ from the center, under a solid angle $\Omega_0$ can be obtained as:

$$P_{\Omega_0}(\omega) = \frac{1}{2} c\varepsilon_0 d^2 \int_0^{\Omega_0} \bar{\bar{W}}_\omega(\vec{r}_1, \vec{r}_1) \, d\Omega \tag{4}$$

where $\vec{r}_1$ are points on the sphere of radius $d$ and $c$ is the speed of light in vacuum.

We consider a SiC rod, 1.62 μm long, with a square cross section of 325X325 nm$^2$ and a Au rod, 3.44 μm long, with the same cross section. The difference in length between the two rods is due to the different effective wavelength for the two materials, corresponding to a λ/2 dipole antenna in vacuum with the desired peak of emission around λ=12 μm. For simplicity all calculations are performed in vacuum. The adopted method allows to consider the system embedded in a homogenous medium provided it is transparent in the wavelength range under examination. Once the material and the cross section are selected, longer rods will emit at longer wavelengths. For both cases we calculated the far field power spectral density collected at $d$=1 mm distance from the antenna over a cone of maximal half-angle θ=15° (numerical aperture NA=sin(θ)=0.26) in forward (left) and backward (right) emission with respect to the antenna main axis (x-axis, as sketched in figure 1) when the antennas are at an equilibrium temperature of 400 K. The calculation has been performed by using cubic volume cells of size ΔV=65x65x65 nm$^3$. In order to check reliability and convergence of our results we performed a calculation on the single rod with a finer discretization. Increasing the number of cells by a factor 8 (ΔV=32.5x32.5x32.5 nm$^3$) the resonance peak shifts of 0.5 % and the relative error on the emitted power is less than 5 %, showing that further discretization is not required.

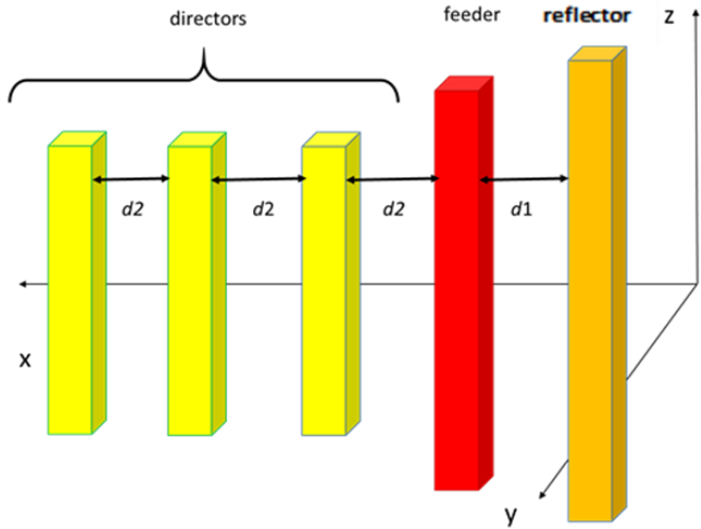

Figure 1: Typical scheme of a 5 elements Yagi-Uda antenna.

In figure 2 we report the normalized power spectral density for both cases in forward emission. Obviously, due to the symmetry of the geometry, the signal collected in forward and backward direction in the case of a single rod is the same. In order to provide a reference, all the quantities have been normalized with respect to the maximum value obtained for the SiC rod. The peak of emission for gold is about five times weaker than for SiC. We also note that there is a single line at 12.02 μm and its full width at half maximum (FWHM) is approximately 100 nm for the SiC antenna. On the other hand, the emission for the gold rod at the same temperature has a much broader spectrum with a FWHM of approximately 2 μm. For both cases the radiative pattern evaluated at the peak of emission has the typical dipole pattern shape. For illustration purposes we report in the inset of Figure (2) the radiation pattern calculated for the SiC antenna at the resonant wavelength of 12.02 μm. The z-axis is the axis of the dipole antenna. For the gold rod the radiation pattern is qualitatively the same.

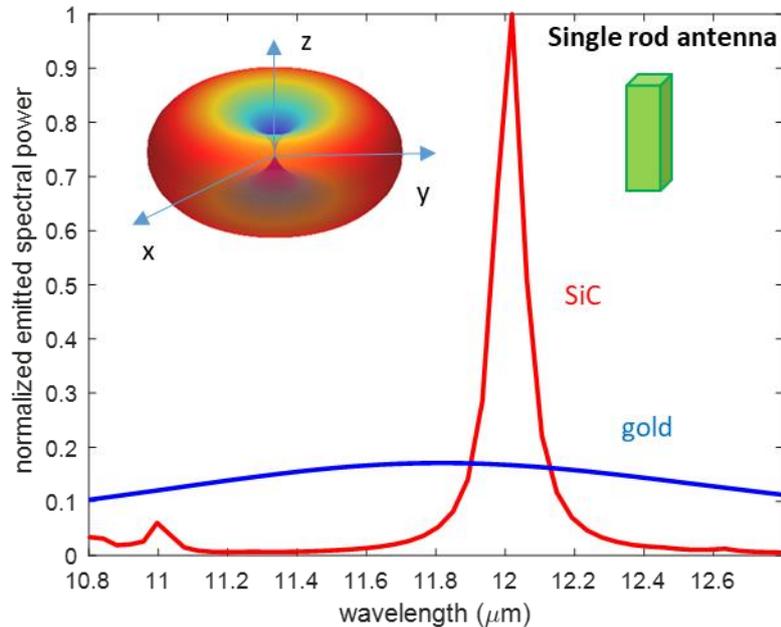

Figure 2: Normalized forward emitted spectral power density collected under a NA=0.26 calculated for the SiC dipole antenna (red) and gold antenna (blue) described in the text. The inset shows the emission pattern at the peak wavelength for the SiC antenna. The emission pattern for the gold antenna is qualitatively the same.

As mentioned earlier, according to standard YU specifications all the director elements are half-wavelength antennas slightly detuned towards shorter wavelengths with respect to the feeder. Similarly, the reflector is detuned to longer wavelengths. We choose gold as material for the rods that will behave as directors and reflector. For simplicity and with the main objective to keep a further realization procedure as easy as possible, we kept the same section for all the elements (325x325 nm$^2$) and selected the resonance wavelength by changing the length of the rods. Following the standard YU criteria [35,36], we started with the ratio between the length of the reflector (director) and the gold feeder equal to 1.15 (0.9). A fine optimization could be performed by changing either the length and the section of these elements but it is beyond the scope of this work. Moreover, our calculations show that small deviations from these values do not significantly affect the final result.

In order to quantify the amount of the background noise introduced by the emission of the not ideal parasitic elements, we firstly evaluated the thermal emissivity of just one reflector element at a temperature T=400 K. We depict in figure 3 the normalized power spectral density emitted by the reflector (dashed line) collected within a cone corresponding to the same NA=0.26 and normalized with respect to the spectral power emitted by the single SiC rod. We note that the reflector does not contribute much to the overall emission in the wavelength range under investigation. On the other hand, performing the same calculations on a set of one, two and three Au directors, separated by a typical 0.3λ distance (according to standard YU specifications [35,36]) we note a very different behavior. Indeed, the single director is by all means a gold antenna tuned at 10.5 μm and its emission is comparable to the gold feeder. Moreover, a system composed by two or more directors is a composite antenna, all the elements are strongly coupled, being separated by a subwavelength distance. Increasing the number of directors, resonant modes at longer wavelengths are created due to the increased size of the system and the resulting amount of "noise" is even higher than the emission level of the SiC feeder with only 3 directors. However, we will show that the SiC feeder narrow emission linewidth is crucial for the boost effect of the total signal from the YU antenna.

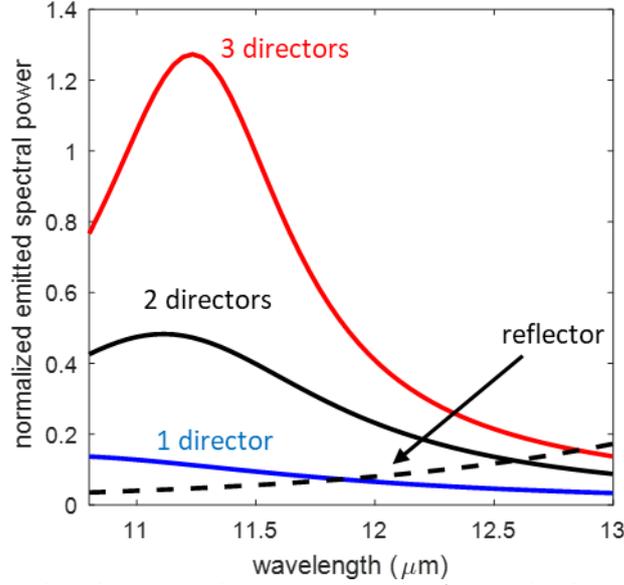

Figure 3: Normalized emitted spectral power density for a single reflector (dashed line), one director ( solid blue line), 2-(solid black) and 3-(solid red) directors at a given temperature of 400 K.

Results

We considered the thermal emission from a 5 elements (reflector, SiC feeder and 3 directors) YU antenna, at an equilibrium temperature of 400 K. We performed a scan to find the optimal distances of reflector and directors, starting from the standard specifications. We considered only two scanning parameters: *d1* being the distance between the feeder and the reflector and *d2* the distance between the feeder and the director, as well as the distance between adjacent directors. In order to find optimized performances, we defined a function of merit taking into account two main objectives: 1) the ratio between the forward ($P_{\Omega_0}(forward)$) and backward ($P_{\Omega_0}(backward)$) emitted power under the same numerical aperture NA=0.26 (function of merit *F1*); 2) the previously investigated gain enhancement with respect to the single SiC rod at the peak wavelength of 12.02 μm (function of merit *F2*). Thus we want to optimize the function:

$$F = F1 * F2 = \frac{P_{\Omega_0}(forward)}{P_{\Omega_0}(backward)} * \frac{P_{\Omega_0}(forward)}{P_{SiC,\Omega_0}(forward)}; \qquad (5)$$

A map of *F* as a function of *d1* and *d2* is depicted in figure (4). We found optimized results by choosing 3.8 μm (0.31λ) as the distance between feeder and director and between each director, while the optimized distance between the feeder and the reflector is 2.4 μm (0.2 λ). With these values the function of merit reaches the maximum of about F=60. We note that performances of the antenna are not strongly affected by small deviations from optimal parameters. Indeed we obtain already F>50 in a wide zone for the parameters *d1* ranging from 1.9-3.1 μm and *d2* from 3.5-4.1 μm.

This feature is extremely interesting in terms of fabrication tolerances of the antenna. In Figure (5) we report the normalized power spectral density collected with a NA of 0.26 in forward emission for both cases of SiC and gold feeder. In the first case we note that despite of the noise introduced by parasitic elements the signal of the feeder is boosted by a factor of 5 and the narrow linewidth (approx. 100 nm) is preserved while the background emission from the directors is partially suppressed.

On the other hand, if the gold feeder is considered, we note that the signal emitted by the feeder is not boosted, the thermal emission is ruled by the directors and the overall effect is an enhancement of the emission of the directors. There is a well-defined emission line of approximately 1 $\mu$m bandwidth tuned at 11.45 $\mu$m over the background noise. These results suggest that the study of a gold thermal antenna must be performed by considering it as an array of coupled emitters/scatterers. The emission peak cannot be evaluated a priori by selecting the resonance of the feeder. In order to put into evidence the different behavior for the case of gold and hybrid antenna we calculated the signal emitted by a system composed only of reflector and directors, without feeder. The result is shown in Figure (5). It is clear that the reflector is too far from the directors to act efficiently and it actually reduces the emission with respect to the system composed by 3 directors only (Figure (3)). Moreover we note that the introduction of the SiC feeder does not modify the contribution from the other elements because it is acting on a very narrow band and it does not efficiently couple with the others. On the other hand, when we add the gold feeder, it behaves as a reflector for the directors, boosting the signal and shifting the wavelength peak due to coupling with the other elements. A finer optimization could be performed for the pure metal antenna nevertheless the large bandwidth related to gold antennas remains the main drawback. Finally we plot the corresponding radiative pattern for both antennas at the peak of emission in figure 6(a-b). We note that in both cases we find the typical directional Yagi Uda emission pattern although the hybrid antenna is more efficient. The radiation patterns for both cases could be object of further optimization. However, this task goes beyond the scope of this work.

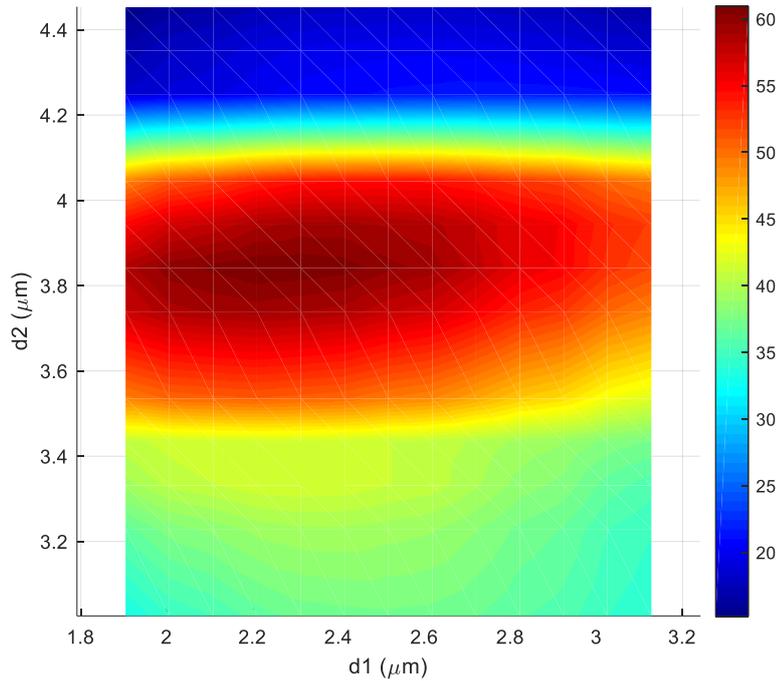

Figure 4: Map of the function of merit F defined in eq. (5) as a function of the scanning parameters *d1* (distance between the feeder and the reflector) and *d2* (distance between the feeder and the director, distance between adjacent directors).

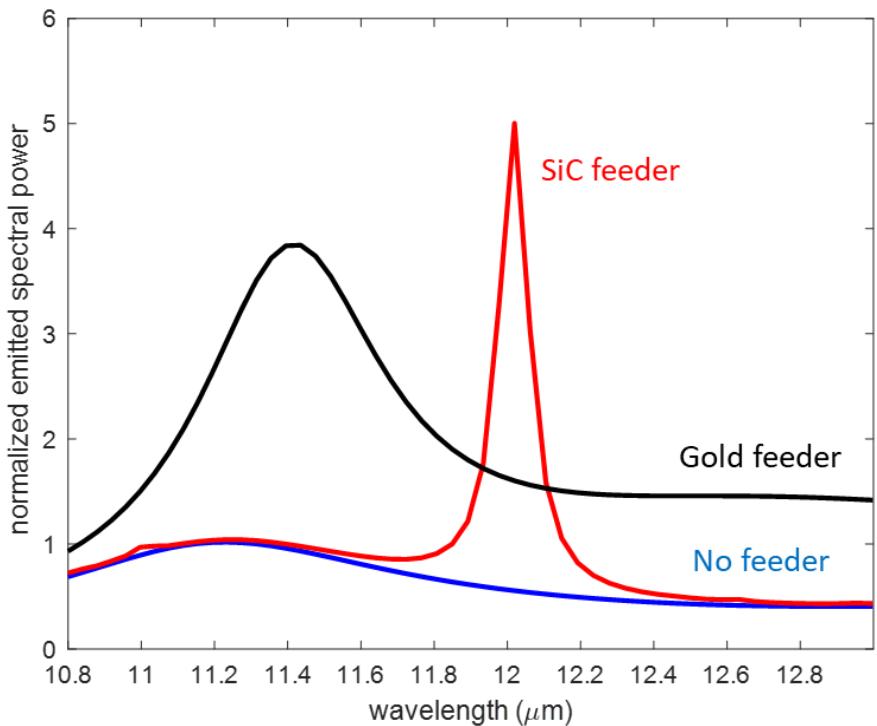

Figure 5: Normalized forward emitted spectral power for a 5-element gold (black) and hybrid gold-SiC feeder (red) Yagi-Uda thermal antenna. The blue line shows the normalized forward emitted spectral power obtained from the parasitic elements only, without any feeder.

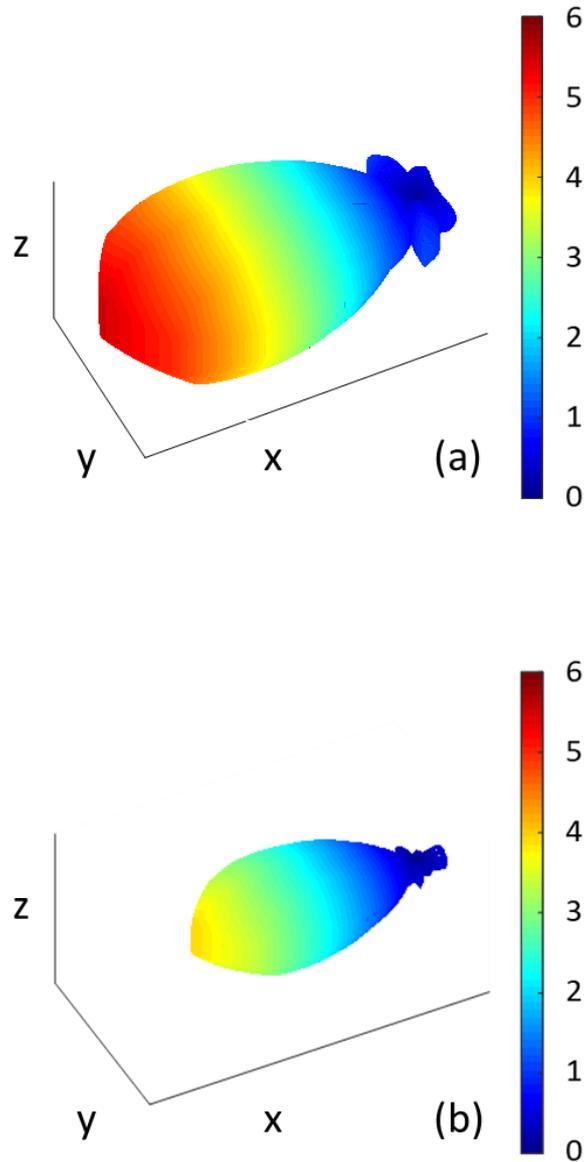

Figure 6: Normalized radiation pattern at the peak wavelength for the hybrid gold with SiC feeder Yagi-Uda antenna (a) and for the gold with gold feeder antenna (b). The quantities are normalized with respect to the maximum emission of a single SiC dipole antenna of the same size of the feeder.

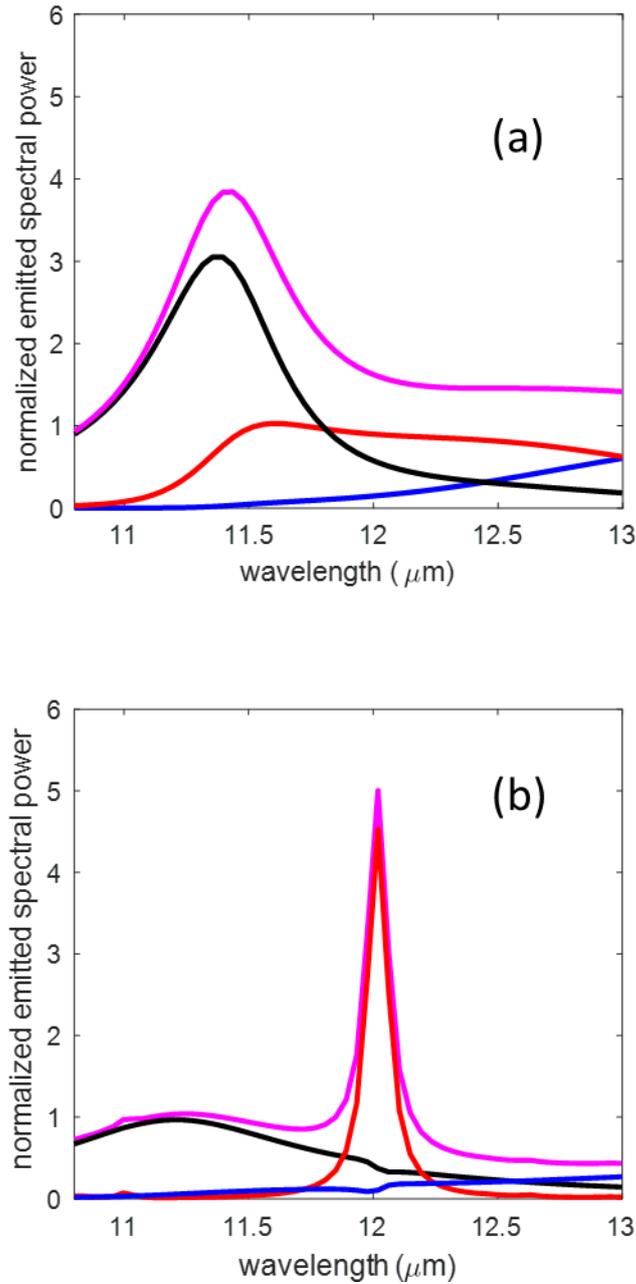

Figure 7: Out of equilibrium emission from the previously studied Yagi-Uda antenna when only the feeder is emitting (red), only the reflector (blue) and only the directors (black) for gold (7a) and hybrid (7b) antenna. As a reference we also show the equilibrium case in which all the elements are at T=400 K (magenta)

     A deeper insight is provided by calculating the out of equilibrium case in which only the feeder or the directors or the reflector are at a given temperature of 400 K while the other elements are not emitting (corresponding to an ideal temperature of 0 K). In this way it is possible to see how each element contributes to the emission in the presence of the others acting only as scatterers by taking advantage of the superposition principle. The results are reported in figure 7 for both metallic and hybrid antennas. In figure 7a it is depicted the result for the gold YU antenna. As expected, the contribution from the directors plays the dominant

role. However, the emission from the feeder is responsible for a shift of the wavelength peak. We also note that, with respect to the emission of the single elements calculated in figure 3 and figure 2 the YU scheme is responsible for the enhancement of emission of every part of the antenna (reflector, feeder and directors). Vice versa, with the hybrid YU (figure 7b), only the emission of the feeder is enhanced giving as a result a higher signal to noise ratio.

Finally we note that the contribution from the thermal emission of the parasitic elements is particularly evident if we want to maximize the contrast between forward and backward emission. In figure 8 it is shown the radiation pattern in the xy-plane at the peak wavelength for the case of the hybrid antenna. We compare the results obtained if only the emission form the feeder is considered (red) with respect to the equilibrium case (black). We note that the backward radiation in the acceptance angle ±15 ° is strongly suppressed if only the feeder is heated. The calculated forward to backward ratio is 25. In the equilibrium case the ratio is reduced to 11. The same argumentation could be performed from the metal YU antenna but in this case the cause of a poor forward to backward ratio is due to the fact that all the elements contribute to the emission and the reflector cannot reflect with the same efficiency the signals coming from elements placed at different distances. In figure 8 we show as a reference the radiation pattern at the peak for the all-gold YU antenna (green) exhibiting a pronounced backward emission.

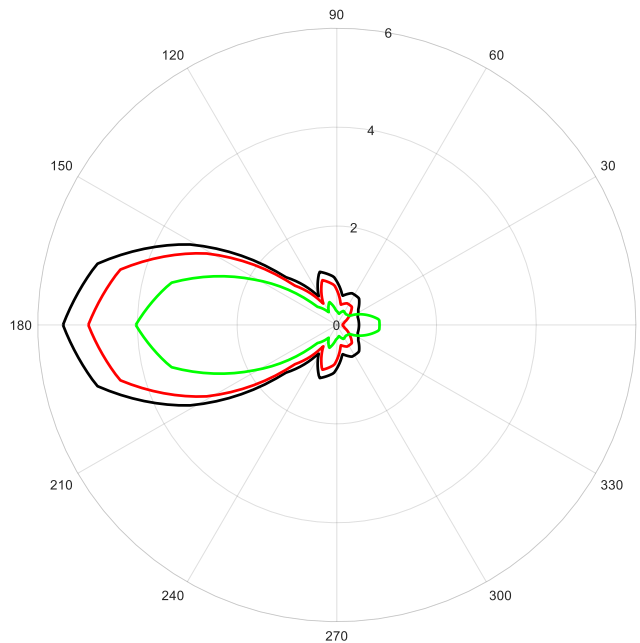

Figure 8: Emission pattern in the x-y plane at the peak wavelength for the hybrid YU antenna at the equilibrium temperature of 400 K (black) and for the out of equilibrium case in which only the feeder is heated (red). We note a reduction of backward emission if the contribution from the parasitic elements is not considered. As a reference we also report the emission pattern at peak wavelength for the gold antenna (green).

As mentioned before, one of the most interesting aspect of YU configuration is its robustness with respect to design parameters. This feature is extremely appealing for easy realization of tunable sources with small wavelength differences and good directionality. As an example we report the emission spectra for slightly different schemes. We considered the same antenna, and the same NA for collecting the emitted signal. We only changed the length of the SiC feeder rod. Beside the previously investigated case with $L_1$= 1.62 μm, we performed calculations for $L_2$=1.36 μm and $L_3$=1.88 μm. From the calculated emission spectra depicted in figure 9 we notice that neither the linewidths nor the efficiencies are affected, there is only a shift in the wavelength of the emission peak. The increased signal for longer rods it is just related to the increasing amount of SiC (see figure 9b) being the calculated power spectral density normalized with respect to the maximum value obtained for the $L_1$ SiC rod, as usual. Arrays of different YU antennas might be used to finely match emission/absorption lines of various hazardous gases and used both as sources and as receivers to create compact low cost, miniaturized sensing devices. Finally we note that the hybrid antenna appears differently with respect to the typical YU antenna (as sketched in figure 9a). Indeed the SiC feeder is smaller than the other elements. This is related to the different effective wavelength for gold and SiC rods.

Despite of the background noise introduced by the not ideal parasitic elements, the coherent emission provided by the SiC antenna feeder makes it possible to boost the signal directivity even with an increased number of directors. However, our calculations show that the best signal to noise ratio is obtained with 3 directors. Depending on the requirements needed for a real application it could be worth developing multiple parameters optimization procedures to evaluate the best trade-off between fabrication costs and performances.

Conclusions:

We show that despite of the chaotic nature of thermal radiation it is possible to obtain efficient highly directional and narrow bandwidth thermal antennas in the mid to far IR by adapting the Yagi-Uda scheme with a combination of metallic and polar materials. In particular, we studied hybrid SiC/Au antennas. Our calculations show that it is possible to tune the emission wavelength by properly choosing the resonance of the feeder and to boost the signal and its directivity. The detrimental effect due to thermal emission of the parasitic elements has been taken into account and we have shown that a pure metallic Yagi-Uda antenna could not efficiently work in the 12 μm range, especially if a narrow spectral linewidth of IR radiation is required. This is due to the fact that enhancement and selectivity of thermal emission in metals is related to resonant plasmon polaritons excitation which is efficient only in a wavelength range close to the plasma frequency of the metal (typically in the visible and near IR). Nevertheless, combining high selectivity of SiC antennas in the 12 μm wavelength range (due to resonant phonon polariton excitation) and high scattering cross section and low losses of metals in that wavelength range efficient hybrid thermal Yagi-Uda antennas can be designed. The enhanced performances and the robustness with respect to design parameters make hybrid thermal Yagi-Uda antennas very promising in both transmission and reception mode for integrated, low cost and mass-market gas sensing devices.

Acknowledgments

M.C. kindly acknowledges Prof. Constantin Simovski for stimulating discussions and suggestions on spatial and temporal coherence of thermal radiation from nanostructured systems. M. C. thanks the CNRS and the group Theory of Light-Matter and Quantum Phenomena of the Laboratoire Charles Coulomb for hospitality during his stay in Montpellier.

M. A. thanks the SAPIENZA University of Rome and the Department of Basic and Applied Sciences for Engineering for hospitality during his stay in Rome under the visiting professor program. M. A. acknowledges support from the Institute Universitaire de France, Paris – France (UE).

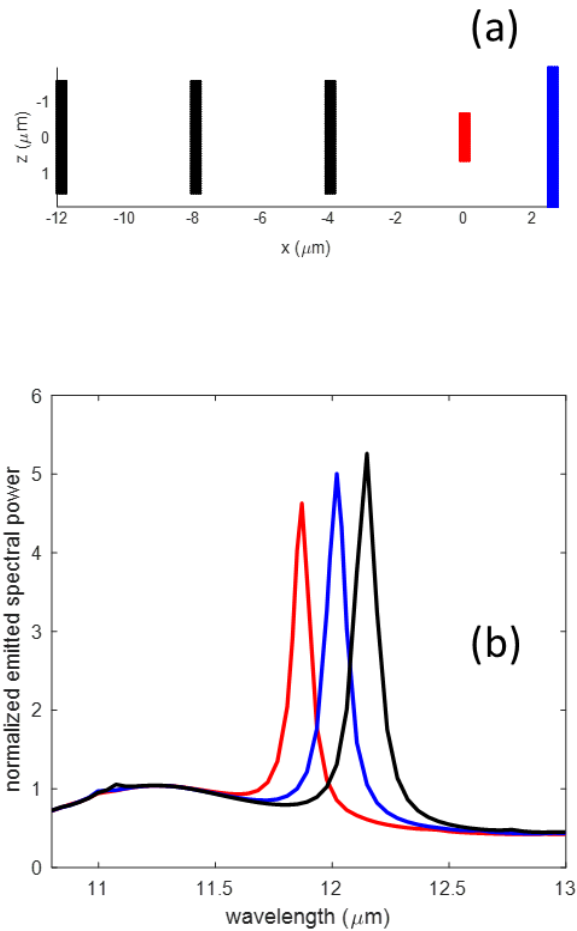

Figure 9: (a) sketch of the hybrid YU antenna with the length of the SiC feeder equal to L2. (b) Normalized emitted spectral power for three hybrid YU antennas obtained by varying the length of the SiC feeder from L1=1.62 μm (blue), L2=1.36 μm (red) and L3=1.88 μm (black)